\documentclass[]{aastex631}

\begin{document}

\title{Tidally Trapped Two-pole Pulsations Discovered in a Close Binary with a Massive $\beta$ Cephei Star}

\author{Ping Li}
\affiliation{Yunnan Observatories, Chinese Academy of Sciences (CAS), 650216 Kunming, China}
\affiliation{University of Chinese Academy of Sciences, No.1 Yanqihu East Rd, Huairou District, Beijing, China 101408}

\author{Wen-Ping Liao$^{*}$}
\affiliation{Yunnan Observatories, Chinese Academy of Sciences (CAS), 650216 Kunming, China}
\affiliation{University of Chinese Academy of Sciences, No.1 Yanqihu East Rd, Huairou District, Beijing, China 101408}

\author{Sheng-Bang Qian}
\affiliation{School of Physics and Astronomy, Yunnan University, Kunming 650091, China}

\author{Li-Ying Zhu}
\affiliation{Yunnan Observatories, Chinese Academy of Sciences (CAS), 650216 Kunming, China}
\affiliation{University of Chinese Academy of Sciences, No.1 Yanqihu East Rd, Huairou District, Beijing, China 101408}

\author{Jia Zhang}
\affiliation{Yunnan Observatories, Chinese Academy of Sciences (CAS), 650216 Kunming, China}

\author{Qi-Bin Sun}
\affiliation{School of Physics and Astronomy, Yunnan University, Kunming 650091, China}

\author{Fang-Bin Meng}
\affiliation{Yunnan Observatories, Chinese Academy of Sciences (CAS), 650216 Kunming, China}
\affiliation{University of Chinese Academy of Sciences, No.1 Yanqihu East Rd, Huairou District, Beijing, China 101408}



\begin{abstract}

Tidally tilted pulsators (TTPs), whose pulsation axis aligns with the binary's semi-major axis, represent a newly established class of oscillators in binary systems. While all previously known TTPs are either $\delta$ Scuti or subdwarf B-type stars, their existence has remained unidentified in more massive $\beta$ Cephei variables. Here, we report the discovery of tidally trapped pulsations in the massive ellipsoidal variable HD~329379, based on photometry from the Transiting Exoplanet Survey Satellite (TESS). Our analysis reveals a $\beta$ Cephei pulsator in a 2.25-day orbit whose pulsation mode amplitude is strongly modulated with the orbital frequency. Based on our analysis, we concluded that this modulation can be explained by pulsations with significantly larger amplitude near the star's two tidal poles (apsides). We interpret this as a tidally distorted quadrupole pulsation chariacteristiced by trapped two-pole pulsations, with a pulsation axis aligned with the tidal axis. This represents the first identification of such a pulsation mode in a $\beta$ Cephei star, which differs from single-sided pulsations observed in previous works, marking a rare and important discovery. Our work extends the family of TTPs beyond $\delta$ Scuti and subdwarf B-type stars to include more massive $\beta$ Cephei variables. In particular, the two-pole pulsator HD~329379 stands out as the prototype of a new class of TTPs in massive stars. These results not only provide a new insight to probe the interior structure and evolutionary state for massive stars but also offer a unique opportunity to study the interaction between pulsations and strong tidal distortions.

\end{abstract}



\section{Introduction} \label{sec:intro}

Asteroseismology, the study of stellar oscillations, provides a powerful tool for probing the internal structure and determining fundamental parameters of stars with high precision \cite{Aerts2010}. This method is significant in astrophysics because our understanding of the universe largely depends on knowledge of the fundamental parameters and structure of stars.

A remarkable demonstration of how internal physical processes can alter pulsations comes from rapidly oscillating Ap (roAp) stars \cite{Kurtz1982}. These objects exhibit high-overtone, nonradial pulsations whose frequency multiplets are split by stellar rotation. The oblique pulsator model \cite{Kurtz1982} explains these observations by proposing that the pulsation axis is aligned with the magnetic axis, which is itself inclined to the rotation axis—making roAp stars the first known class of pulsators with a misaligned pulsation axis. Subsequent refinements suggest the pulsation axis lies within the plane defined by the magnetic and rotational axis \cite{Bigot2011}. A classic manifestation is the distorted quadrupole mode, observed in many roAp stars \cite{Balona2011, Holdsworth2015, Holdsworth2016, Holdsworth2017, Holdsworth2018a, Holdsworth2018b, Holdsworth2019, Shi2020, Shi2021}, which exhibits strong amplitude modulation (with primary and secondary peaks at rotational phases 0.0 and 0.5) and a continuous phase shift throughout the rotation cycle.

In close binary systems, tidal forces can play a role analogous to magnetic fields in roAp stars, by tilting the pulsation axis. A striking example is the tidally tilted pulsators (TTPs) that are an intriguing new class of oscillating stars in binary systems \cite{Handler2020, Fuller2020}; In such stars, the pulsation axis coincides with the line of apsides, or semi-major axis, of the binary. This misalignment can lead to a clear amplitude modulation of nonradial pulsations throughout the orbital cycle. A defining characteristic of TTPs in edge-on systems (the orbital inclination $i \approx 90^\circ$) is that the observer views the pulsating star across a full range of latitudinal angles, which greatly facilitates the identification of pulsation modes \cite{Jayaraman2022}.

TTPs were first conclusively discovered in both the $\delta$ Scuti ($\delta$ Sct) and subdwarf B-type (sdB) stars using photometric data from the Transiting Exoplanet Survey Satellite (TESS; \cite{Ricker2015}). The first unambiguous detection of tidally tilted pulsations came from the close binary HD~74423, a $\delta$ Sct star with an 8.8\,d$^{-1}$ mode whose amplitude and phase modulate over its 1.58\,d orbit—consistent with tidal confinement toward the $L_1$ or $L_3$ side \cite{Handler2020}. Subsequent studies have identified TTPs in other systems: CO Cam hosts at least four 14\,d$^{-1}$ modes localized near $L_1$ \cite{Kurtz2020}; TIC~63328020 has a 21\,d$^{-1}$ mode consistent with a distorted $\ell=1$, $|m|=1$ mode \cite{Rappaport2021}; and HD~265435 shows 27 orbital-frequency-split multiplets in an sdB-dwarf binary \cite{Jayaraman2022}. 

The prevalence of misaligned pulsations in roAp stars provides a crucial precedent for understanding TTPs. The tidally trapped pulsators (a subset of TTPs, as seen in HD~74423 and CO~Cam) demonstrate the extreme case of this misalignment, with pulsations largely confined to a single hemisphere. This naturally leads to the question of whether an intermediate configuration exists: could some tidally trapped pulsators exhibit two-sided pulsations, with significant amplitudes on both sides or near the two apesides?

However, no such frequency pattern characteristic of two-sided pulsations, analogous to the distorted multiplet patterns seen in roAp stars but caused by tidal distortion, has been identified in binary systems. This constitutes one key gap in our observational knowledge. Furthermore, all known TTPs have been found exclusively in $\delta$ Sct or sdB stars. Despite theoretical works \cite{Fuller2020, Fuller2025} suggesting their possibility, tidally tilted pulsations have not been detected in the more massive $\beta$ Cep variables, representing another significant observational absence. We therefore first conducted a systematic search for two-sided pulsators in binary systems containing massive $\beta$ Cep variables.

HD~329379 (TIC 122314621) emerges as a promising candidate to fill this gap. This object is a close B-type binary with a visual magnitude of 9.83 mag \cite{Zacharias2012}, but there is otherwise limited available literature. It was initially classified as a pulsating ellipsoidal variable \cite{Steindl2021}, and later suggested to host $\beta$ Cep-type pulsations \cite{Eze2024}. ref. \cite{Steindl2021} also reported pulsation frequency splitting modulated by the orbital period in this system, suggesting possible origins such as tidally trapped, tidally perturbed, or tidally induced pulsations—including heartbeat effects.

\section*{Observations and corresponding analysis}

\begin{figure}
	\centering
	\includegraphics[width=\linewidth]{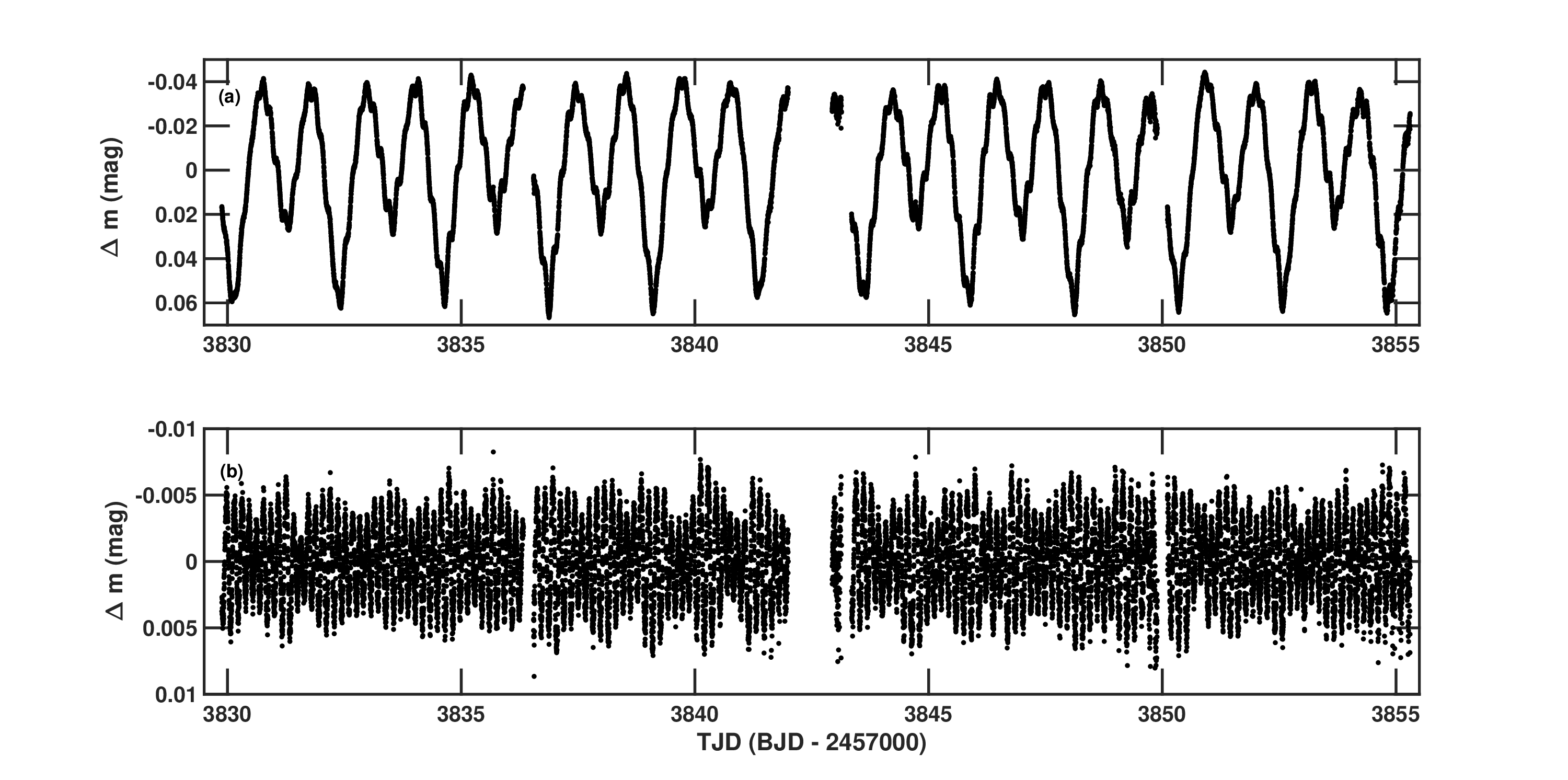}
	\caption{%
		\textbf{TESS light curves of HD~329379.}
		\textbf{(a)}, A segment of the light curve displaying both ellipsoidal variations and high-frequency pulsations. Similar features are present in the other observed sector (not shown).
		\textbf{(b)}, The same segment after removal of the orbital variations and low-frequency artifacts, revealing the modulation of the pulsation amplitude with the orbital period. Time is given in TESS Julian Date (TJD = BJD $-$ 2,457,000), where BJD is the Barycentric Julian Date.}
	\label{fig:Lightkurve}
\end{figure}

TESS recently observed the close binary HD~329379 in Sectors 66 and 93 at a two-minute cadence, providing an opportunity to investigate the system's pulsations. Fig.~\ref{fig:Lightkurve} shows the two-minute cadence light curve of Sector 93 of HD~329379. Panel (a) shows a full light curve of the sector. The full light curves from two sectors are too compressed at this scale to reveal the detailed structure. Similar to HD~74423 \cite{Handler2020}, the light curve of HD~329379 exhibits minima of alternating depth with relatively constant maxima, a characteristic signature of ellipsoidal variables \cite{Morris1985} rather than rotational variability as previously proposed by ref.  \cite{Bernhard2015}. This alternating depth arises in part from differential gravity darkening near the $L_1$ and $L_2$ Lagrangian points of the tidally distorted star. After the orbital variability has been removed, Fig.~\ref{fig:Lightkurve}(b) clearly shows that the pulsation amplitude is modulated with the orbital period.

\begin{figure}
	\centering
	\includegraphics[width=1.0\linewidth]{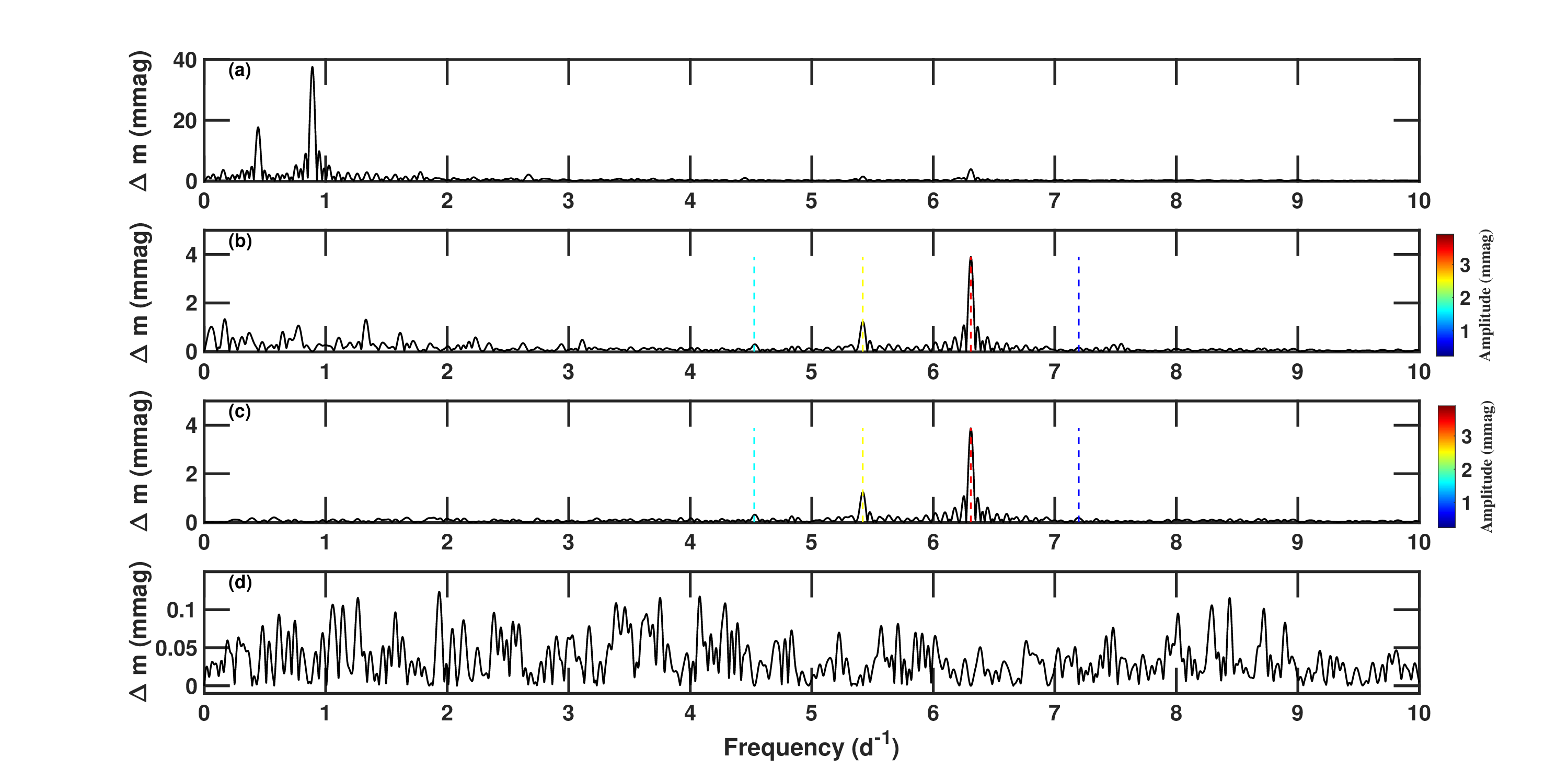}
	\caption{%
		\textbf{Fourier amplitude spectra of the TESS light curve for HD~329379.}
		\textbf{(a)}, Initial amplitude spectrum. The highest peak corresponds to twice the orbital frequency ($2\nu_{\mathrm{orb}}$; $\nu_{\mathrm{orb}} = 0.445665 \pm 0.000009\,\mathrm{d}^{-1}$), characteristic of ellipsoidal variation.
		\textbf{(b)}, Amplitude spectrum of the residuals after pre-whitening with a sixteen-term harmonic series at the orbital frequency. The pulsation multiplet centered on $\nu_1 = 6.309599 \pm 0.000009\,\mathrm{d}^{-1}$ (see Table~\ref{tab:frequency_multiplets}) is visible, along with low-frequency instrumental artifacts.
		\textbf{(c)}, Same as panel (b) after applying a high-pass filter to remove the low-frequency artifacts, clearly revealing the multiplet structure. 
		\textbf{(d)}, Residual amplitude spectrum after subtracting the multiplet mode detailed in Table~\ref{tab:frequency_multiplets}. Note the change in scale.
	}
	\label{fig:Spectrum}
\end{figure}

The initial amplitude spectrum are shown in Fig.~\ref{fig:Spectrum} (a). As expected for ellipsoidal variables with double-wave light curves, the highest peak corresponds to the second harmonic of the orbital frequency. The derived orbital frequency is $\nu_{\mathrm{orb}} = 0.445665 \pm 0.000009\,\mathrm{d}^{-1}$, with corresponding orbital periods $P_{\mathrm{orb}}=1/\nu_{\mathrm{orb}} = 2.24358 \pm 0.00005$ d. Using the \textit{O-C} method (see Methods), we obtained a corrected orbital period of $2.24630 \pm 0.00006$ d.  
\begin{deluxetable*}{l c c c c l c c c c}
	\tabletypesize{\scriptsize}
	\label{tab:frequency_multiplets}
	\tablecaption{Frequency multiplets for $\nu_1$ in sectors 66 (left) and 93 (right). Numbers in parentheses indicate 1$\sigma$ uncertainties in the last digits.}
	\tablehead{
		\colhead{ID} & \colhead{Freq. (d$^{-1}$)} & \colhead{Ampl. (mmag)} & \colhead{Phase (rad)} & \colhead{S/N} & 
		\colhead{ID} & \colhead{Freq. (d$^{-1}$)} & \colhead{Ampl. (mmag)} & \colhead{Phase (rad)} & \colhead{S/N}
	} 
	\startdata
	$\nu_1 - 4\nu_{\mathrm{orb}}$ & 4.530003 (80) & 0.308 (1) & 0.738 (22) & -- & 
	$\nu_1 - 4\nu_{\mathrm{orb}}$ & 4.526883 (65) & 0.246 (4) & 0.147 (62) & -- \\
	$\nu_1 - 2\nu_{\mathrm{orb}}$ & 5.418601 (12) & 1.371 & 0.590 (8) & -- & 
	$\nu_1 - 2\nu_{\mathrm{orb}}$ & 5.419595 (20) & 1.329 & 0.896 (6) & -- \\
	$\nu_1$ & 6.309599 (9) & 4.037 & 0.453 (2) & -- & 
	$\nu_1$ & 6.309136 (7) & 3.906 & 0.876 (21) & -- \\
	$\nu_1 + 2\nu_{\mathrm{orb}}$ & 7.201456 (95) & 0.250 & 0.099 (19) & -- & 
	$\nu_1 + 2\nu_{\mathrm{orb}}$ & 7.196821 (83) & 0.242 & 0.473 (138) & -- \\
	\enddata
\end{deluxetable*}

Fig.~\ref{fig:Spectrum}(b) shows the amplitude spectrum of the residuals after subtracting a sixteen-harmonic fit of the orbital variation at frequency $\nu_{\mathrm{orb}}$. The pulsation multiplet is clearly visible, along with low-frequency instrumental artefacts. These artefacts were removed using a high-pass filter, resulting in the cleaned amplitude spectrum shown in Fig.~\ref{fig:Spectrum} (c), which displays only the pulsation frequency and its orbital sidelobes.

Fig.~\ref{fig:Spectrum}(c) reveals that most of the pulsational variation is concentrated in a frequency multiplet centred on the highest amplitude peak. We extracted the frequencies within this multiplet and examined their separations using Period04 software \cite{Lenz2005}, adopting a frequency resolution of $\delta\nu = 1 / \Delta T = 0.037\,\mathrm{d}^{-1}$, where $\Delta T$ is the continuous observing baseline of one TESS sector. All extracted frequencies are separated, to within the frequency resolution, by the orbital frequency $\nu_{\mathrm{orb}} = 0.445665 \pm 0.000009\,\mathrm{d}^{-1}$. This confirms that the structure consists of a single pulsation frequency at $\nu_{\mathrm{1}} = 6.309599 \pm 0.000009\,\mathrm{d}^{-1}$ and its orbital sidelobes.

We therefore employed a combined linear and nonlinear least-squares optimization approach in Period04 to refine the frequencies, amplitudes, and phases of the multiplet components. The results of this analysis are provided in Table~\ref{tab:frequency_multiplets}. Uncertainties in frequency, amplitude, and phase were estimated following the method of ref. \cite{Montgomery1999}. The frequency ratio of the pulsation to the orbital frequency is $14.1644226 \pm 0.0000287$, which deviates from the nearest integer (14) by approximately $5730\sigma$; this strongly suggests that the pulsation is not tidally excited.

\begin{figure}
	\centering
	\includegraphics[width=\linewidth]{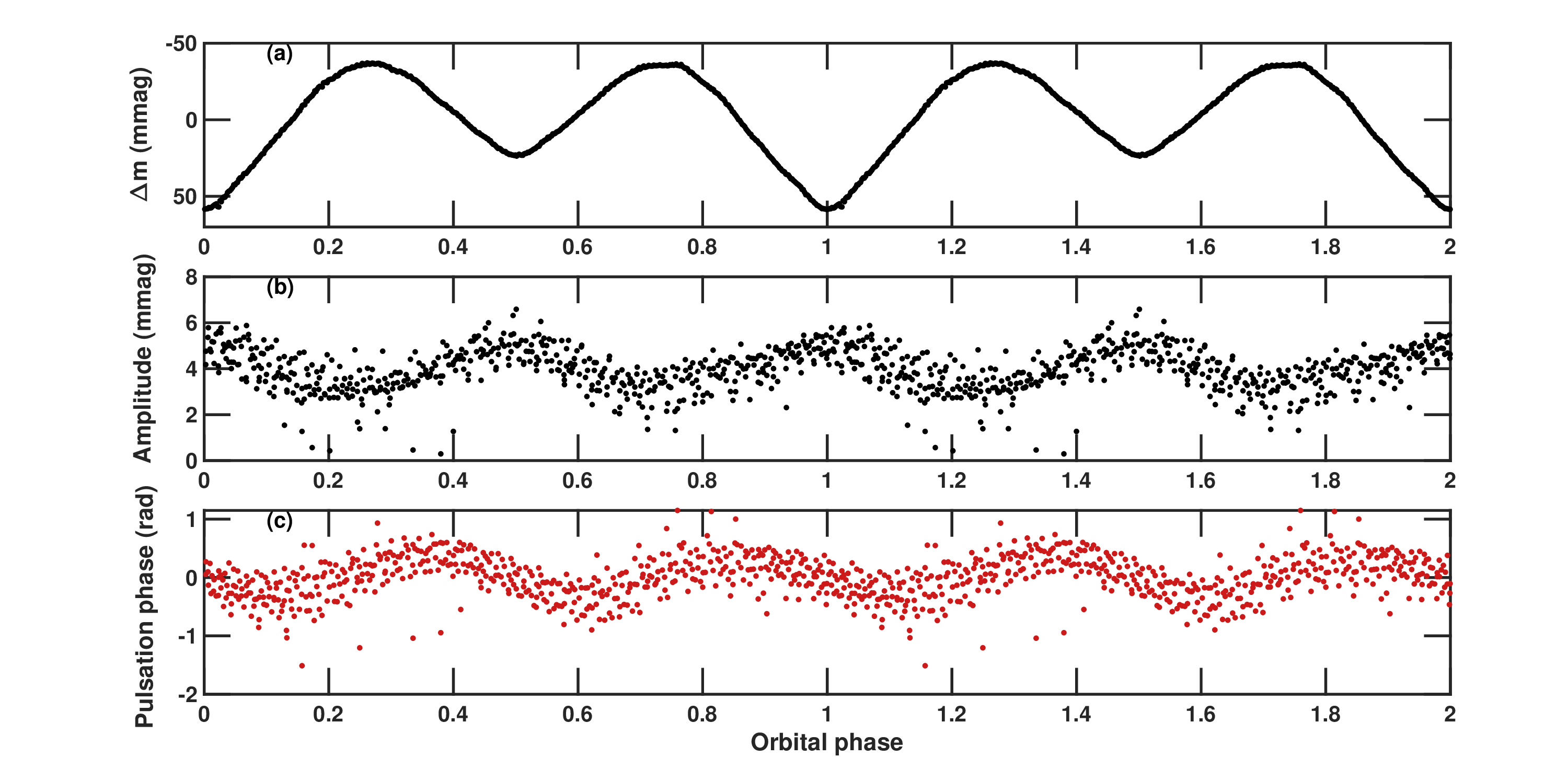} 
	\caption{\textbf{Amplitude and phase modulation of the pulsation over the orbital period for HD~329379.}
		\textbf{(a)}, Orbital light variations are shown as a function of orbital phase for comparison. The data have been binned to 0.002 in phase and the pulsation variations removed.
		\textbf{(b)}, Pulsation amplitude variations (black dots) at $\nu_1 = 6.309599~\mathrm{d}^{-1}$ versus orbital phase. The zero time point is set to $t_0 = \mathrm{BJD}\,2460100.07681$, which was chosen so that the first two orbital sidelobes have an equal phase. As expected in the oblique pulsator model, it can be seen that the orbital light minimum coincides with the pulsation maximum.
		\textbf{(c)}, Variations in pulsation phase (red dots) over the orbital cycle of HD~329379. A systematic phase reversal occurs periodically.
	}
	\label{fig:Lightcurve}
\end{figure}

Finally, adopting the method from ref. \cite{Handler2020}, we performed least-squares fits of the pulsation frequency $\nu_1$ for HD~329379 to 0.1 d segments of the data to investigate the amplitude and phase variations throughout the orbital cycle. The original light curve was binned into phase intervals of 0.002, and the resulting curves for the system are shown in Fig.~\ref{fig:Lightcurve}. This figure reveals that the orbital light minimum coincides with the pulsation maximum at the time when the line of sight aligns with the line of apsides.

We model the orbital variation of pulsation amplitude and phase using the same method of magnetically modulated variability in roAp stars \cite{Unno1979} as in the TTP HD 74423 study \cite{Handler2020}. The luminosity variations are assumed to originate from temperature fluctuations $\delta T$ on the stellar surface, expressed as a decomposition in spherical harmonics:
\begin{equation}\label{eq:temperature_perturbation}
	\delta T \propto e^{i\omega t} \sum_{\ell=0}^{2} A_{\ell} Y_{\ell}^{0} (\theta_p, \phi_p)
\end{equation}
where $i$ is the imaginary unit, $\omega$ is the pulsation frequency, $A_{\ell}$ is the amplitude of the mode, and $Y_{\ell}^{0}$ denotes the spherical harmonic function. The coordinates $(\theta_p, \phi_p)$ are defined in a frame whose axis aligns with the line of apsides, implying that the pulsation is axisymmetric with respect to the tidal axis. Transforming these coordinates first to a frame aligned with the rotation axis, and then to the inertial frame oriented along the line of sight, followed by integration over the visible hemisphere at each epoch, yields the luminosity variation as a function of time:
\begin{equation}\label{eq:luminosity_variation}
	\Delta L(t) \propto e^{i\omega t} \sum_{\ell=0}^{2} N_{\ell} A_{\ell} \sum_{m=-\ell}^{\ell} d_{m,0}^{\ell} (\beta) \, d_{0,m}^{\ell} (i_0) \, e^{-i m \Omega t}
\end{equation}
Here, $m$ is the spherical azimuthal order of the pulsation mode, $N_{\ell}$ is a normalization factor defined in equation (23) of ref. \cite{Saio2004}, and the coefficients $d_{m,0}^{\ell}(\beta)$ and $d_{0,m}^{\ell}(i_0)$ arise from the coordinate transformations: from $(\theta_p, \phi_p)$ to the rotation-aligned frame (where the pulsation axis is inclined by $\beta$ relative to the rotation axis), and then to the observer’s frame (where the line of sight is inclined by an angle $i_0$ relative to the rotation axis). (For further details, see refs. \cite{Unno1979} and \cite{Saio2004}).
\begin{figure}[ht]
	\centering
	\includegraphics[width=1\linewidth]{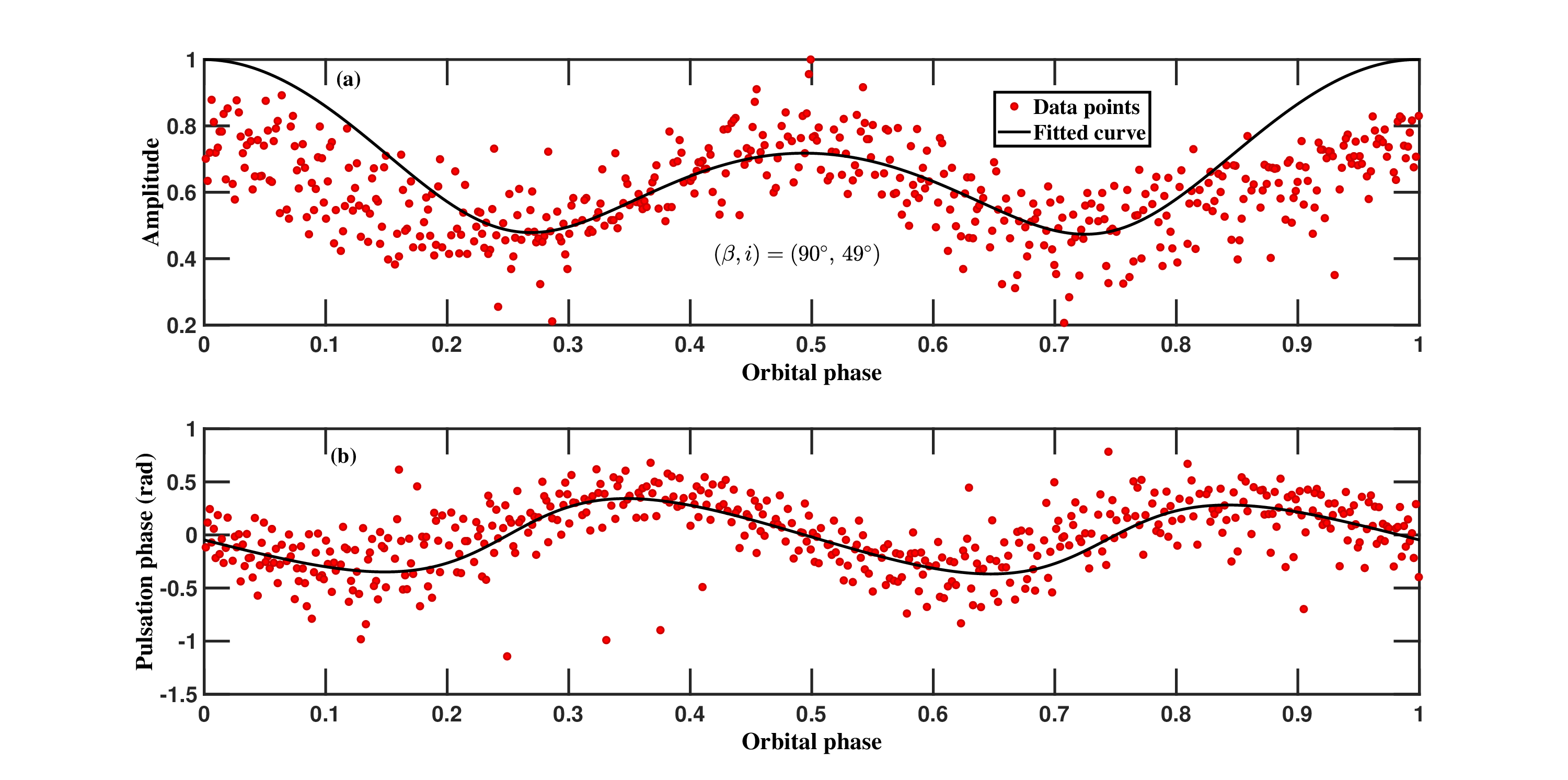} 
	\caption{\textbf{Comparison of observed amplitude and phase modulation with the oblique pulsator model.}
		\textbf{(a)}, The amplitude modulation of HD~329379 (red dots) is shown alongside the model (black lines).
		\textbf{(b)}, The same comparison, but for phase modulation. 
	}
	\label{fig:tidal_modulation} 
\end{figure}

We assumed a pulsation axis lying in the orbital plane with $\beta = 90^\circ$ and adopted an inclination angle of $i_0 = 49^\circ$ from our binary modelling (see Methods, Table~\ref{tab:WD}). We then searched for a set of complex amplitude ratios $(A_1, A_2)/A_0$ that best reproduce the observed amplitude and phase modulation of HD\,329379. The best-fit model yields $A_1/A_0 = -0.192747 - 0.041040i$ and $A_2/A_0 = 13.590981 + 0.926956i$. A comparison between the model and observations is shown in Fig.~\ref{fig:tidal_modulation}. The model reproduces the overall amplitude and phase variability reasonably well, although discrepancies in amplitude are present around orbital phases 0.1 and 0.9. The amplitude deviations between the model and the data cannot be reconciled within an axisymmetric eigenfunction assumption with respect to the pulsation axis, indicating that the pulsation is not perfectly aligned with the tidal axis. 

Our model reveals that the tidally tilted pulsations in HD~329379 are dominated by a tidally distorted quadrupole mode. This mode involves the coupling of radial, dipole, and quadrupole components, with its pulsation axis aligned with the tidal axis (the line of apsides) due to the tidal bulge created by its binary companion. According to the theoretical framework in ref. \cite{Fuller2025}, the radial mode, the $Y_{10x}$ mode, and the $Y_{22+}$ mode couple in a strong tidal environment where the $l_t\ge$ 3 components of the tidal potential dominate. This cross-$l$ multi-mode synergy results in tidally trapped pulsations. This manifests as follows: at orbital apsides (phases 0.0 and 0.5), the line of sight is aligned with the apsides, yielding two symmetric luminosity peaks with continuous phase variation (Fig.~\ref{fig:tidal_modulation}). At quadrature (phases 0.25 and 0.75), the near-perpendicular orientation of the tidal axis to the line of sight reduces the amplitude, but the modulation remains smooth. Consequently, the pulsation amplitude is predominantly concentrated near the two tidal poles, resulting in two-pole pulsations and the observed periodic phase variations. The modelled surface amplitude distribution of the pulsation is shown in Supplementary Fig.~\ref{fig:Temp_per}, which again clearly illustrates the higher amplitude near the two tidal poles.

\section*{Summary and conclusions}
The pulsating component in the binary system HD~329379 exhibits tidally distorted quadrupole pulsations, which are caused by the tidal bulge raised by its companion. This two-sided pulsation mode produces a frequency multiplet split by the orbital frequency. The resulting frequency and phase patterns demonstrate that the pulsation axis is aligned with the line of apsides. Notably, HD~329379 is the first known oblique  $\beta$ Cep pulsator in which the pulsation axis is governed by tidal distortion. More importantly, the pulsations are predominantly confined to the two tidal poles—a spatial distribution that differs markedly from that of single-side pulsators, such as HD 74423 \cite{Handler2020} and CO Cam \cite{Kurtz2020}. This distinct pattern suggests a new tidally tilted mode characterized by trapped two-pole pulsations.  Although the identity of the pulsating component in HD~329379 remains unconfirmed, this does not affect the conclusions of our analysis. Future spectroscopic observations using the Southern African Large Telescope (SALT) could resolve this ambiguity, given the star's location in the southern hemisphere.

The pulsation mode observed in HD~329379 is both rare and significant for a $\beta$ Cep star, holding considerable value for asteroseismology. However, there must be a class of stars that includes $\beta$ Cep and other pulsating variables which pulsate in tidally distorted quadrupole modes, display two-pole pulsations, and have their pulsation axes aligned with their tidal axes. This discovery provides impetus for searching for more similar stars. For instance, many pulsating stars in eclipsing systems that have been reported in ref. \cite{Eze2024} could be investigated further to search for potential new TTPs in $\beta$ Cep. HD~329379 inspires more detailed studies of the interaction between stellar pulsations and tidal distortion in massive binary stars. Tidally tilted pulsations of sdB stars have been used to investigate their internal structure and evolution (see, for example, ref. \cite{Jayaraman2022}). This suggests that detecting and analysing tidally tilted pulsations among $\beta$ Cep stars could provide new insights into the evolution of massive stars in close binary systems and enable us to probe their interior structure.

\begin{acknowledgments}
This work is supported by the Science Foundation of Yunnan Province (Grant Nos. 202503AP140013, 202501AS070055, 202401AS070046), the International Partnership Program of Chinese Academy of Sciences (No. 020GJHZ2023030GC), the 2022 CAS “Light of West China” Program, the National Natural Science Foundation of China (No. 12573038 and 12503040), the Postdoctoral Fellowship Program of CPSF under Grant Number GZC20252095, the China Postdoctoral Science Foundation under Grant Number 2025M773194, and the Caiyun Postdoctoral Program in Yunnan Province of China (grant No. C615300504124). This work has made use of data from the European Space Agency (ESA) mission Gaia. (\href{https://www.cosmos.esa.int/gaia}{https: //www.cosmos.esa.int/gaia}), processed by the Gaia Data Processing and Analysis Consortium (DPAC; \href{https://www.cosmos.esa.int/web /gaia/dpac/consortium}{https://www.cosmos.esa.int/web /gaia/dpac/consortium}). Funding for the DPAC has been provided by national institutions, in particular the institutions participating in the Gaia Multilateral Agreement. The TESS data presented in this paper were obtained from the Mikulski Archive for Space Telescopes (MAST) at the Space Telescope Science Institute (STScI). STScI is operated by the Association of Universities for Research in Astronomy, Inc. Support to MAST for these data is provided by the NASA Office of Space Science. Funding for the TESS mission is provided by the NASA Explorer Program.
\end{acknowledgments}

%






\bibliography{sample631}{}
\bibliographystyle{aasjournal}



\end{document}